\documentclass{PoS}

\title{Dynamical Simulations with Highly Improved Staggered Quarks}

\ShortTitle{Dynamical Simulations with Highly Improved Staggered Quarks}

\author{K.~Y.~Wong\\
        Department of Physics \& Astronomy,
        University of Glasgow, Glasgow, UK G12 8QQ\\
	E-mail: \email{k.wong@physics.gla.ac.uk}}

\author{\speaker{R.~M.~Woloshyn}\\
        TRIUMF, 4004 Wesbrook Mall, Vancouver, British Columbia,
        Canada V6T 2A3\\
        E-mail: \email{rwww@triumf.ca}}

\abstract{It is well established that lattice artifacts can be suppressed
substantially by the use of SU(3)-projected smeared links in the fermion
action. An example is the Highly Improved Staggered Quark action where the
ASQ-like effective links are constructed from reunitarized Fat7 links.
A general procedure is presented for computing the derivative of the fermion
action with respect to the base links (fermion force) --- a key component in
dynamical simulations using molecular dynamics evolution. The method is
iterative and can be applied to actions with arbitrary levels of smearing
and reunitarization. The cost of calculating the fermion force is determined
for the ASQ action and the HISQ action. Test results show that
calculating the HISQ force is about two times more expensive than the ASQ
force.}

\FullConference{The XXV International Symposium on Lattice Field Theory\\
                 July 30 - August 4 2007\\
                 Regensburg, Germany}

\begin{document}

\section{Introduction}

The staggered quark formalism provides a fast method to simulate light
quarks at small quark masses. Staggered fermions, however, have the
property that each lattice quark field describes four identical
quark flavors (``tastes'') rather than one, and coupling to gauge
fields leads to taste-changing errors, which are $O(a^{2})$
effects~\cite{Lepage99}. Taste-changing interactions can be suppressed
substantially by the use of fat
links~\cite{Lepage99,Kostas99a,Lagae99,Kostas99b};
accurate simulation results~\cite{Davies04,Aubin04,Kronfeld06} were obtained
using the ASQTAD action, where the base links are replaced by Fat7
effective links (see Fig.~1). The success of the ASQTAD
action suggests that further improvement may be possible with more smearing.
Perturbative calculations and non-perturbative results of
Refs.~\cite{Follana03,Follana07a} show that
significant improvement can be obtained by first projecting the fat links
back to SU(3) matrices before being used in the next fattening process.
The result is the Highly Improved Staggered Quark action (HISQ) where the
ASQ-like effective links are constructed from reunitarized Fat7 links
(another widely used smearing technique is HYP-smearing~\cite{Hasenfratz01},
which is based on SU(3)-projected fat links also)
\begin{equation}
U^{eff}(U^{R},U^{R\dag}), \quad
U^{R}(U^{F},U^{F\dag}), \quad
U^{F}(U,U^{\dag}).
\end{equation}
Here $U$ is the base link, $U^{F}$ is the Fat7 link, $U^{R}$
is the reunitarized fat-link and $U^{eff}$ is the ASQ-like effective
link (see Fig.~1). The HISQ action has no tree-level order $a^{2}$ errors
like the ASQTAD action,
and has taste-changing interactions that are 3-4 times smaller than the
ASQTAD action. In addition, the action has no tree-level order $(am)^{4}$
errors to leading order in the quark's velocity $v/c$, it therefore provides
an accurate discretization of the charm quark on the lattice. For example this
action has been used recently to obtain high precision results for $D$ meson
and decay constants for $\pi$, $K$, $D$ and
$D_{s}$~\cite{Follana07a,Follana07b}.

So far all unquenched simulations using staggered quarks as light quarks
were done with the ASQTAD action. Given the nice features of
the HISQ action, it is desirable to use it for the sea quarks also.
Dynamical simulations with HISQ, however, are complicated by the extra
level of fattening and SU(3)-projection. In this paper we present a general
procedure for computing the derivative of the fermion action with respect
to the base links (fermion force) --- a key component in dynamical
simulations using molecular dynamics evolution. This method is
iterative and can be applied to actions with arbitrary levels of
smearing or SU(3)-projection. We compare the efficiency of the algorithm
for the ASQTAD and HISQ actions on small lattices.

Our treatment of unitarized links have been influenced by
Kamleh {\it et al.}~\cite{Kamleh04}. Other approaches have been
discussed in Refs.~\cite{Morningstar04,DeGrand05,Hasenfratz07}.
\begin{figure}
\centering
\includegraphics[width=0.6\textwidth]{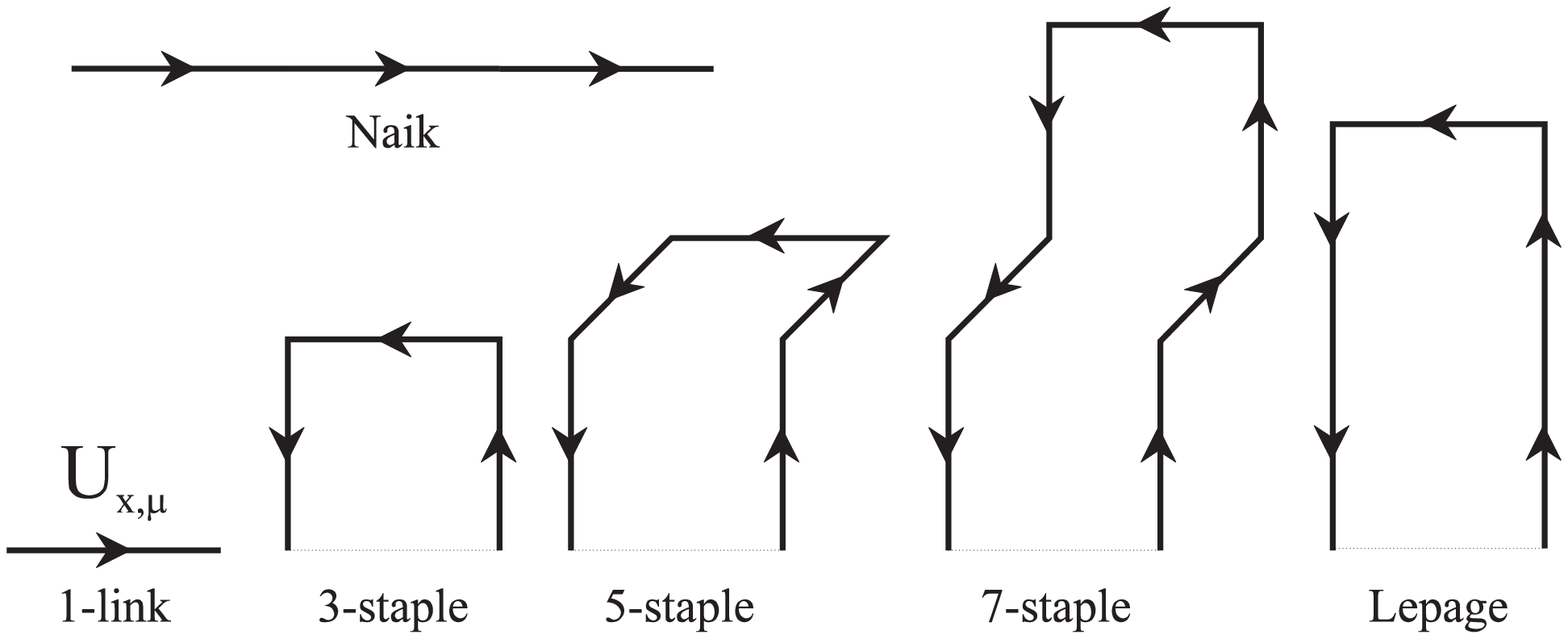}
\caption{Paths used in the ASQTAD and HISQ actions. Path coefficients
for ASQTAD can be found in Ref.~\cite{Kostas99b}. The HISQ effective
links are constructed by first applying a Fat7 fattening to the
base links ($U\to U^{F}$ with coefficients 1-link:$1/8$, 3-staple:$1/16$,
5-staple:$1/64$, 7-staple:$1/384$), then a SU(3)-projection ($U^{F}\to U^{R}$),
and finally an ``ASQ'' smearing ($U^{R}\to U^{eff}$ with coefficients
1-link:$1+\epsilon/8$, 3-staple:$1/16$, 5-staple:$1/64$, 7-staple:$1/384$,
Lepage:$-1/8$, Naik:$-(1+\epsilon)/24$, where the parameter $\epsilon$
is introduced to remove $(am)^4$ errors~\cite{Follana07a}).}
\end{figure}

\section{The Fermion Force}

The staggered quark action is (we follow the notations of~\cite{Gottlieb87})
\begin{equation}
S_{f} = \left< \Phi\left|\left[
M^{\dag}[U]M[U]
\right]^{-n_{f}/4} \right|\Phi \right>,
\end{equation}
where
\begin{eqnarray}
M_{x,y}\left[U\right]
& = &
2m\delta_{x,y}+D_{x,y}\left[U\right]
\nonumber \\
& = &
2m\delta_{x,y}+\sum_{\mu}\eta_{x,\mu}
\left( U_{x,\mu}^{eff}\delta_{x,y-\mu}-U^{eff\dag}_{x-\mu,\mu}\delta_{x,y+\mu}
\right).
\end{eqnarray}
The pseudo-fermion field $\Phi$ is defined on even lattice sites
only to avoid a doubling of flavors from using $M^{\dag}M$ instead of $M$
in the action. This procedure is valid since $M^{\dag}M$ has no
matrix element connecting even and odd lattice sites.

A key component in dynamical simulations using molecular dynamics
evolution is the computation of fermion force --- derivative
of the fermion action with respect to the base links
\begin{equation}
f_{x,\mu} = \frac{\partial S_{f}}{\partial U_{x,\mu}}
= \frac{\partial}{\partial U_{x,\mu}}
\left<\Phi\left|\left[M^{\dag}[U]M[U]\right]^{-n_{f}/4}\right|\Phi\right>.
\end{equation}
The derivative can be computed straightforwardly if $n_{f}$ is
a multiple of 4; for other numbers of fermion flavors
the 4th-root of $M^{\dag}M$ can be approximated by a rational
expansion (the RHMC algorithm~\cite{Kennedy99,Clark03,Clark06})
\begin{equation}
[M^{\dag}M]^{-n_{f}/4} \approx \alpha_{0}
+ \sum_{l} \frac{\alpha_{l}}{M^{\dag}M+\beta_{l}},
\end{equation}
where $\alpha_{l}$ and $\beta_{l}$ are constants. The derivative becomes
{\setlength\arraycolsep{2pt}
\begin{eqnarray}
\frac{\partial S_{f}}{\partial U_{x,\mu}}
& = &
-\sum_{l} \alpha_{l} \left<\Phi[M^{\dag}M+\beta_{l}]^{-1}\left|
\frac{\partial}{\partial U_{x,\mu}}\left(M^{\dag}[U]M[U]\right)
\right|[M^{\dag}M+\beta_{l}]^{-1}\Phi\right>
\nonumber \\
& = &
-\sum_{l} \alpha_{l} \left(
\left< X^{l} \left|
\frac{\partial D^{\dag}[U]}{\partial U_{x,\mu}}
\right| Y^{l} \right>
+
\left< Y^{l} \left|
\frac{\partial D[U]}{\partial U_{x,\mu}}
\right| X^{l} \right>
\right),
\end{eqnarray}}
with $|X^{l}\rangle=[M^{\dag}M+\beta_{l}]^{-1}|\Phi\rangle$ and
$|Y^{l}\rangle=D|X^{l}\rangle$. Note that $X^{l}$ and $Y^{l}$ are defined
on even and odd sites respectively. Taking the derivatives of $D$, $D^{\dag}$
with respect to $U^{eff}$, $U^{eff\dag}$ and writing out the matrix indices
we have
\begin{equation}
\left[ f_{x,\mu} \right]_{ab}
=
\frac{\partial S_{f}}{\partial \left[U_{x,\mu}\right]_{ab}}
=
\sum_{y,\nu} (-1)^{y} \eta_{y,\nu}
\left( \frac{\partial [U^{eff}_{y,\nu}]_{mn}}{\partial [U_{x,\mu}]_{ab}}
\left[f^{(0)}_{y,\nu}\right]_{mn}
+
\frac{\partial [U^{eff\dag}_{y,\nu}]_{mn}}{\partial [U_{x,\mu}]_{ab}}
\left[f^{(0)\dag}_{y,\nu}\right]_{mn} \right),
\end{equation}
where $f^{(0)}_{y,\nu}$ is the vector outer product
of the field variables at $y$ and $y+\nu$
\begin{equation}
\left[f^{(0)}_{y,\nu}\right]_{mn}= \left\{
\begin{array}{ll}
{\displaystyle
\sum_{l} \alpha_{l} [Y^{l}_{y+\nu}]_{n} [X^{l*}_{y}]_{m}
}
&
\quad \textrm{for even $y$}
\vspace{0.1cm} \\
{\displaystyle
\sum_{l} \alpha_{l} [X^{l}_{y+\nu}]_{n} [Y^{l*}_{y}]_{m}
}
&
\quad \textrm{for odd $y$}
\end{array}\right.,
\end{equation}
and the sum on $y$, $\nu$ extends over the effective links that
contain the base link $U_{x,\mu}$, and $(-1)^{y}=1$ for even $y$'s and
$(-1)^{y}=-1$ for odd sites. Note that the force is equal to
$f^{(0)}$ when there is no smearing, i.e., $U^{eff}=U$. Furthermore
all complications associated with the rational expansion have been
absorbed into $f^{(0)}$ and the derivatives $\partial U^{eff}/\partial U$,
$\partial U^{eff\dag}/\partial U$ are calculated only once.

\section{Smearing and SU(3)-Projection}

For actions with multiple levels of smearing or SU(3)-projection,
such as the HISQ action, we use the chain rule to compute
$\partial U^{eff}/\partial U$, $\partial U^{eff\dag}/\partial U$
{\setlength\arraycolsep{2pt}
\begin{eqnarray}
\left[ f_{x,\mu} \right]_{ab}
& = &
\sum_{y,\nu;z,\rho} (-1)^{y} \eta_{y,\nu} \Bigg\{
\left( \frac{\partial [U^{eff}_{y,\nu}]_{mn}}{\partial [U^{R}_{z,\rho}]_{pq}}
\frac{\partial [U^{R}_{z,\rho}]_{pq}}{\partial [U_{x,\mu}]_{ab}}
+
\frac{\partial [U^{eff}_{y,\nu}]_{mn}}{\partial [U^{R\dag}_{z,\rho}]_{pq}}
\frac{\partial [U^{R\dag}_{z,\rho}]_{pq}}{\partial [U_{x,\mu}]_{ab}}
\right)
\left[f^{(0)}_{y,\nu}\right]_{mn}
\nonumber \\
& & \hspace{2.4cm}
+ \Bigg( U^{eff} \to U^{eff\dag} \Bigg)
\left[f^{(0)\dag}_{y,\nu}\right]_{mn} \Bigg\}
\nonumber \\
& = &
\sum_{z,\rho}
\left( \frac{\partial [U^{R}_{z,\rho}]_{pq}}{\partial [U_{x,\mu}]_{ab}}
\left[f^{(1)}_{z,\rho}\right]_{pq}
+
\frac{\partial [U^{R\dag}_{z,\rho}]_{pq}}{\partial [U_{x,\mu}]_{ab}}
\left[f^{(1)\dag}_{z,\rho}\right]_{pq} \right),
\end{eqnarray}}
where the sum on $z$, $\rho$ extends over the reunitarized
links $U^{R}$ that contain the base link $U_{x,\mu}$, and
\begin{equation}
\left[f^{(1)}_{z,\rho}\right]_{pq} = \sum_{y,\nu} (-1)^{y} \eta_{y,\nu}
\left( \frac{\partial [U^{eff}_{y,\nu}]_{mn}}{\partial [U^{R}_{z,\rho}]_{pq}}
\left[f^{(0)}_{y,\nu}\right]_{mn}
+
\frac{\partial [U^{eff\dag}_{y,\nu}]_{mn}}{\partial [U^{R}_{z,\rho}]_{pq}}
\left[f^{(0)\dag}_{y,\nu}\right]_{mn} \right).
\end{equation}
The expression for $f_{x,\mu}$ still contains the composite derivative
$\partial U^{R}/\partial U$. Repeat this step until all the derivatives
are explicit
\begin{equation}
\begin{array}{c}
{\displaystyle
\left[f^{(2)}_{z,\rho}\right]_{rs} =
\frac{\partial [U^{R}_{z,\rho}]_{pq}}{\partial [U^{F}_{z,\rho}]_{rs}}
\left[f^{(1)}_{z,\rho}\right]_{pq}
+
\frac{\partial [U^{R\dag}_{z,\rho}]_{pq}}{\partial [U^{F}_{z,\rho}]_{rs}}
\left[f^{(1)\dag}_{z,\rho}\right]_{pq}
},
\vspace{0.2cm} \\
{\displaystyle
\left[ f_{x,\mu} \right]_{ab}=
\left[f^{(3)}_{x,\mu}\right]_{ab} = \sum_{z,\rho}
\left( \frac{\partial [U^{F}_{z,\rho}]_{rs}}{\partial [U_{x,\mu}]_{ab}}
\left[f^{(2)}_{z,\rho}\right]_{rs}
+
\frac{\partial [U^{F\dag}_{z,\rho}]_{rs}}{\partial [U_{x,\mu}]_{ab}}
\left[f^{(2)\dag}_{z,\rho}\right]_{rs} \right)
}.
\end{array}
\end{equation}
Note that $f^{(2)}$ is local since $U^{R}_{z,\rho}$ is a function of
$U^{F}_{z,\rho}$, $U^{F\dag}_{z,\rho}$ only.
Therefore to construct the fermion force  one starts with $f^{(0)}$,
computes $f^{(i)}$ for each smearing and SU(3)-projection working
in towards the first level of smearing. This procedure is very
general and can be applied to actions with arbitrary levels of smearing
and reunitarization.

Two types of derivatives are involved: derivatives
of smeared links ($\partial U^{eff}/\partial U^{R}$ and
$\partial U^{F}/\partial U$) and derivatives of
reunitarized links ($\partial U^{R}/\partial U^{F}$). Implementing the
smeared links and their derivatives is relatively straightforward.
For SU(3)-projected smeared links we use polar decomposition since it is
differentiable and therefore suitable for dynamical simulations
(another popular choice is the stout link introduced by
Morningstar and Peardon~\cite{Morningstar04})
\begin{equation}
{U}^{R}= \frac{\tilde{U}^{R}}{(\mathrm{det}\tilde{U}^{R})^{1/3}},
\quad
\tilde{U}^{R}=U^{F}\left[U^{F\dag} U^{F}\right]^{-1/2}.
\end{equation}
There are several ways to compute the derivative
$\partial U^{R}/\partial U^{F}$. Refs.~\cite{Morningstar04,Hasenfratz07}
utilize the Cayley-Hamilton Theorem; in Ref.~\cite{Follana07a} the
derivative is obtained by solving a matrix equation. Here we follow
Ref.~\cite{Kamleh04} by adopting a rational expansion for
$\left[U^{F\dag} U^{F}\right]^{-1/2}$
\begin{equation}
\left[U^{F\dag} U^{F}\right]^{-1/2} = c_{0}
+ \sum_{l} \frac{c_{l}}{U^{F\dag} U^{F} + d_{l}},
\end{equation}
where $c_{l}$ and $d_{l}$ are constants. A nice feature of the rational
approximation is that the derivatives of $U^{R}$ are calculated easily
{\setlength\arraycolsep{2pt}
\begin{eqnarray}
\frac{\partial [\tilde{U}^{R}]_{pq}}{\partial [U^{F}]_{rs}} & = &
\delta_{pr} \left[ c_{0}
+ \sum_{l}\frac{c_{l}}{U^{F\dag}U^{F}+d_{l}} \right]_{sq}
\nonumber \\
& - &
\sum_{l} c_{l} \left[ U^{F} \frac{1}{U^{F\dag}U^{F}+d_{l}}
U^{F\dag} \right]_{pr}
\left[ \frac{1}{U^{F\dag}U^{F}+d_{l}} \right]_{sq},
\end{eqnarray}
}
and including the determinant
\begin{equation}
\frac{\partial [U^{R}]_{pq}}{\partial [U^{F}]_{rs}}
= (\mathrm{det} \tilde{U}^{R})^{-1/3} \left\{
-\frac{1}{3} \mathrm{tr}\left( \tilde{U}^{R^{-1}}
\frac{\partial \tilde{U}^{R}}{\partial [U^{F}]_{rs}} \right)
[\tilde{U}^{R}]_{pq}
+ \frac{\partial [\tilde{U}^{R}]_{pq}}{\partial [U^{F}]_{rs}}
\right\},
\end{equation}
where the trace is taken with respect to the indices on $\tilde{U}^{R}$.
Finally the derivative can also be computed numerically
\begin{equation}
\frac{\partial [U^{R}]_{pq}}{\partial [U^{F}]_{rs}}
=
\frac{[U^{R}]_{pq} \left( [U^{F}]_{rs}+h \right)
- [U^{R}]_{pq} \left( [U^{F}]_{rs}-h \right)}{2h},
\end{equation}
with $U^{F\dag}$ being held fixed. The matrices $U^{R}(U^{F}\pm h)$
can be obtained by diagonalizing $U^{F}U^{F\dag}$. This method, however,
is relatively inefficient compared to the rational approximation approach
(typically by a factor of five). On the other hand, it provides a
useful check on our approximate calculation.

\section{Benchmark}

We test the performance of the HISQ force on small lattices
with a scalar code. We first compare the average plaquettes obtained
using the RHMC algorithm and the R-algorithm. A high acceptance rate
provides an excellent check of the code. Results are shown in Fig.~2. 
\begin{figure}
\centering
\includegraphics[width=0.6\textwidth]{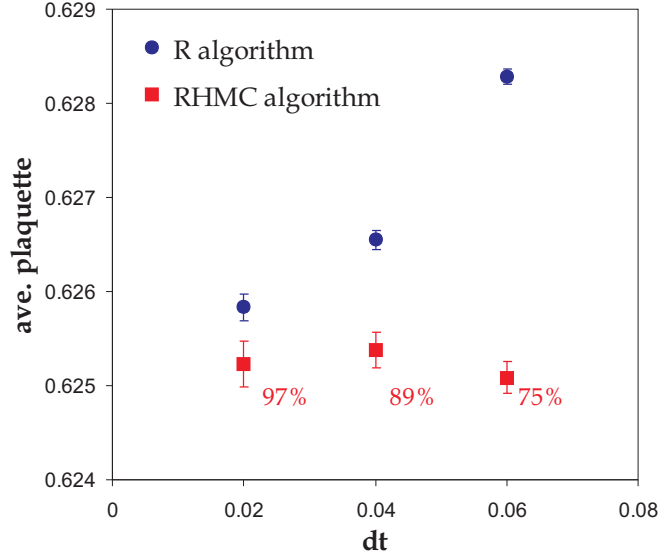}
\caption{Average plaquettes obtained using the RHMC algorithm and the
R-algorithm for the HISQ action (with $\epsilon=1$). Simulation
parameters are $\beta_{pl}=8.0$, $n_{f}=2$, $am=0.25$, $V=4^{4}$ and
$n_{md}\times dt\sim O(1)$ where $dt=0.02,0.04,0.06$ is the
molecular dynamics step size and $n_{md}=50,25,16$ is the number of steps.
The 1-loop Symanzik-improved gluon action is used with tadpole factor
$u_{0}=0.8897$. The RHMC acceptance rates are shown in red.}
\end{figure}

Fig.~3a compares the cost of different components for the ASQTAD action
and the HISQ action at different lattice volumes. Results show that matrix
inversions (with no optimization) dominate for ASQTAD while the computation
of fermion force and inversions are comparable for HISQ. It is interesting
to note that the cost of calculating $f^{(2)}$, derivatives of
reunitarized links, is relatively small compared to $f^{(1)}$
and $f^{(3)}$, derivatives of smeared links. This implies that the HISQ
force is only about twice as expensive as the ASQ force. This point is
further emphasized in Fig.~3b where the ratio of the cost of fermion force to
the cost of gauge force is plotted for the two actions. Finally, the
insensitivity of our results to lattice volume suggests that Fig.~3 is
valid for larger lattices also.
\begin{figure}
\centering
\begin{minipage}{0.505\textwidth}
\includegraphics[width=\textwidth]{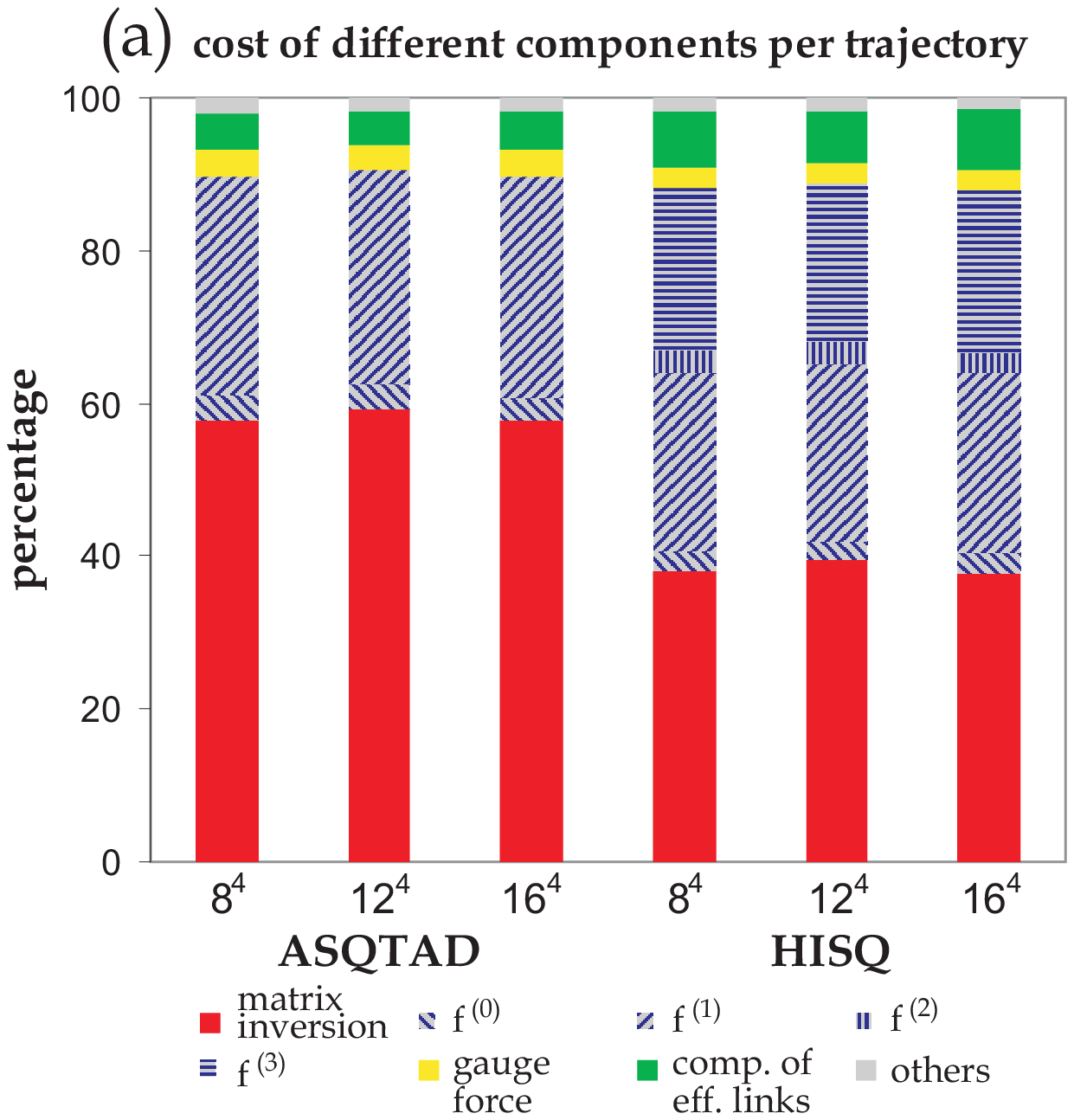}
\end{minipage}%
\begin{minipage}{0.495\textwidth}
\includegraphics[width=\textwidth]{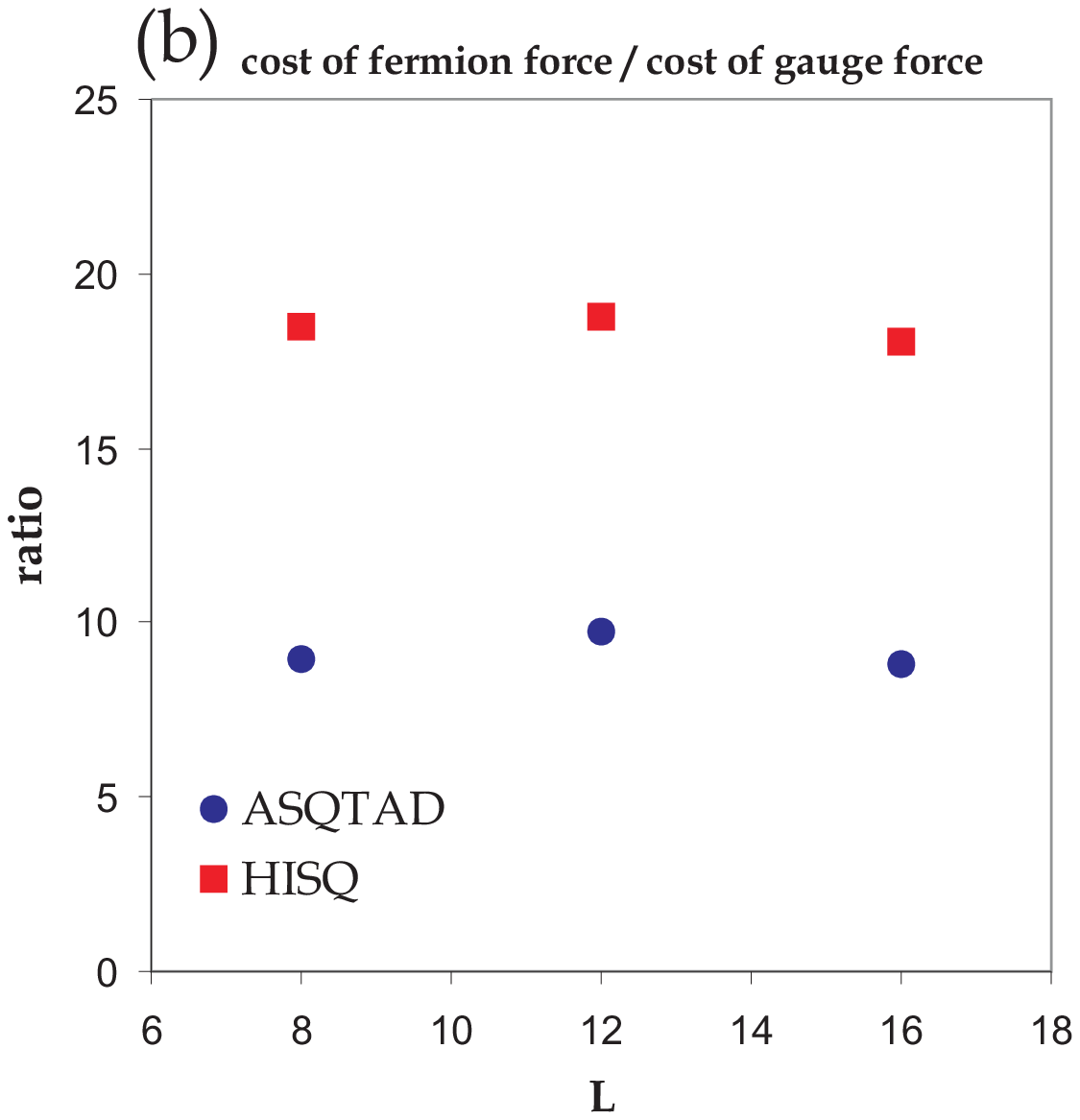}
\end{minipage}
\caption{a) Cost of different components: $f^{(0)}$ is the force when there
is no smearing, $f^{(1)}$ and $f^{(3)}$ are the derivatives of smeared
links and $f^{(2)}$ is the derivative of reunitarized links.
Simulation parameters are the same as those in Fig.~2 and the RHMC
algorithm is used. b) Ratio of the cost of fermion force to the cost of
gauge force for ASQTAD and HISQ.}
\end{figure}

\section{Conclusion}

A general procedure is presented for computing the fermion force for
actions constructed with SU(3)-projected smeared links.
Application to the HISQ action has been discussed and tests have been
done on small lattices. Our results show that the HISQ force is only
two times more expensive than the ASQ force, with most of the additional
cost attributed to the extra level of Fat7-smearing. Given that it is
relatively inexpensive to compute the fermion force one should seriously
consider using SU(3)-projected smeared links in future simulations.

\section{Acknowledgments}

We are very grateful to all our collaborators, and particularly Kent
Hornbostel, for many useful discussions. Computations were done on
facilities provided by WestGrid. This work was supported in part by
the Natural Science and Engineering Research Council of Canada, and
the Particle Physics and Astronomy Research Council of the UK.

\end{document}